# PHASE TRANSITIONS IN THE ISM — A SOURCE OF DISSIPATIVE BEHAVIOUR

A. C. Quillen[1,2] & C. B. Quillen[3]

March 15, 1995

[1] E-mail: quillen@payne.mps.ohio-state.edu
[2] Ohio State University, Department of Astronomy, 174 West 18th Avenue, Columbus, OH 43210
[3] Division of Applied Mathematics, Brown University, Providence, RI 02912

## ABSTRACT

In this paper, by considering a simple fluid model, we investigate the role of phase transitions in the ISM on the galaxy scale gas dynamics. Cooling and heating timescales in the ISM are typically shorter than typical galactic rotation timescales, so the individual phases in the ISM can be assumed to be in temperature equilibrium with the radiation field. Using this approximation we can construct an equation of state which depends upon the average density and mass fractions in the individual phases. Previous studies suggest that there is an equilibrium phase fraction as a function of pressure. We incorporate evolution towards this equilibrium state as a relaxation term with a time to obtain equilibrium $\tau$. We derive a condition in terms of a critical Mach number when one dimensional shocks should be continuous. For small values of the relaxation time $\tau$ we show that the relaxation term acts like a viscosity. We show with one dimensional simulations that increasing $\tau$ causes shocks to become smoother. This work suggests that phase changes can strongly effect the gas dynamics of the ISM across spiral arms and bars.

*Subject headings:* ISM: – ISM: kinematics and dynamics – ISM: shocks

## 1. INTRODUCTION

Recent studies of spiral galaxies have found that the column density of HI across spiral arms (e. g. in M51, Tilanus 1990), peaks at a different location when compared to the $H_2$ (traced in CO) column density (Rand & Tilanus 1990). These observations, coupled with the recent work of Elmegreen (1993) who modeled the dependence of the molecular to atomic mass fraction in clouds as a function of pressure and the UV radiation field, suggest that molecular hydrogen can be disassociated and produced by pressure and radiation field changes across a spiral arm or galactic bar. In this paper, by considering a simple fluid model, we investigate the role of phase transitions in the ISM on the gas dynamics. Our long term goal is to construct multiphase models of the ISM that can be tested through observations of gas at different temperatures and densities in spiral and barred galaxies where changes in the pressure and radiation field are expected as a function of position in the galaxy.

The interstellar medium (ISM) of the Galaxy is inhomogeneous consisting of clouds of cold (temperature $T \lesssim 10^2$ K) atomic and molecular gas, regions of warm ($T \sim 10^4$ K) gas, both neutral and ionized, and regions of hot ($t \sim 10^6$ K) gas. Since the sound crossing time is smaller than other timescales such as the cooling timescale, it is generally assumed that the pressure in each phases is equal to the mean interstellar pressure, except in the molecular phase where self-gravity becomes important. For recent reviews see McKee (1990), Kulkarni & Heiles (1988), Scoville & Sanders (1987), and Begelman (1990). The first theoretical study of a multiphase ISM was the two phase model (Spitzer 1958, Field 1962, Field, Goldsmith & Habing 1969). Later, based on X-ray observations, this model was extended to include third phase by Cox & Smith (1974), and McKee & Ostriker (1977) proposed that the hot phase is maintained by supernovae explosions. More recent studies include details of the heating and cooling balance (Shull & Woods 1985), the role of conductivity and cloud evaporation (Begelman & McKee 1990, McKee & Begelman 1990), pattern formation in conduction fronts (Elphick, Regev, & Shaviv 1992, Aharonson, Regev, & Shaviv 1994), and the interactions of stars and a multiphase medium in a galactic fountain (Rosen, Bregman, & Norman 1993, Chiang & Prendergast 1985). Most of these models have assumed that the average pressure in the galactic plane is static. However, the theories of spiral density waves and gas response in barred galaxies (e. g. Roberts, Shu & van Albada 1979) predict pressure variations as a function of position in a galaxy. This provides an observational setting were we can explore the effect of pressure changes in the ISM.

In this paper we consider the effect of phase changes on the gas dynamics of the ISM. In §2 using the assumption that a equilibrium phase mix exists at a given pressure and density we state a simple set of continuum fluid equations which incorporate transition towards equilibrium as a relaxation term. The characteristics in this dynamical system are given, and a criterion in terms of a critical Mach number is given for when shocks should be continuous. In §3 a set of one dimensional simulations are presented. It is shown from these simulations that increasing the relaxation time causes the shocks to become more smooth. For a small relaxation time, $\tau$, we show that the relaxation term acts like a viscosity. A short discussion follows in §4.

## 2. Continuum Equations

In §2.1 we show that cooling timescales are generally faster than rotational timescales in a galaxy. This justifies using a first approximation where each phase is considered to be in thermal equilibrium with the radiation field. This approximation is particularly desirable since dynamical equations to describe the thermal state of the fluid in each phase are not required (see Pistinner & Shaviv 1993 for a more general set of multiphase fluid equations). We can then use continuum fluid equations where the variables represent averages of local quantities over size scales larger than individual clouds, and far fewer equations are required to describe the multiphase fluid. In §2.2 we describe a simple set of equations to describe such a system. We model the evolution of the phase mix in terms of evolution towards an equilibrium state with $\tau$, the relaxation timescale or timescale required to obtain equilibrium. In §2.3 in the one-dimensional system sound speeds are given for two limits of the system, $\tau \to 0$ and $\tau \to \infty$. To gain insight on this set of equations in §2.4 we derive a dispersion relation for perturbations from equilibrium. We show that for small $\tau$, the relaxation term causes damping similar to that caused by viscosity.

### 2.1 Timescales

We estimate the cooling timescales in order to compare these timescales to galactic timescales. In the cold gas $T \sim 100°$K, Pottasch, Wesselius & van Duinen (1979) from observations of the CII 158$\mu$ line estimate that the cooling rate per Hydrogen nucleon is $1.2 \times 10^{-25}$ erg/s. This implies a cooling timescale of $t_{cool} \sim kT/(dE/dt) \sim 10^4$ years which is fast compared to galactic scales. Likewise, in the ionized medium with $T \sim 8000°$ K and the cooling for the same pressure is about the same (Kulkarni & Heiles 1988, Shull & Woods 1985) so that kT/(dE/dt)$\sim 4 \times 10^5$ years. For comparison, one galactic rotation period in a galaxy with circular rotation speed $v_c$ at a radius $r$ is

$$T = \frac{2\pi r}{v_c} \sim 3 \times 10^7 \text{yr} \left(\frac{200 \text{km s}^{-1}}{v_c}\right) \left(\frac{r}{1 \text{kpc}}\right) \tag{2.1.1}$$

Variations in pressure and density caused by a spiral density wave or barred potential should extend over a solid angle which corresponds to a time that is larger than few percent of the rotation period. Therefore, except in the inner regions of galaxies (e. g. in nuclear rings, where the rotation timescale is faster) heating and cooling timescales are generally faster than the gas response to a barred potential or a spiral density wave. This implies that in each phase, the state of the gas can be described as a function of the radiation field and the average pressure. We note that a warm ($T \sim 8000°$ K) neutral component is estimated to have a much longer cooling time $\sim 10^7$ years (see references in Kulkarni & Heiles 1988) and so may not be in equilibrium with the radiation field. In this paper we do not consider the role of this component of the ISM.

### 2.2 Continuum Fluid Equations

Using the approximation that each phase is in thermal equilibrium with the radiation field, we do not require equations to describe the thermal state of the fluid in each phase. We adopt here a formalism where the variables are averages over size scales larger than the size of individual clouds. We assume here that the mean velocity $v$ of all phases is the same so that all phases move together. Mass conservation is then given by

$$\frac{\partial \rho}{\partial t} + \nabla \cdot (\rho \mathbf{v}) = 0 \tag{2.2.1}$$

where $\rho$ is the average density summed over a scale greater than that of individual clouds. Conservation of momentum or the Euler's equation is

$$\frac{\partial \mathbf{v}}{\partial t} + (\mathbf{v} \cdot \nabla)\mathbf{v} = -\frac{\nabla P}{\rho} \tag{2.2.2}$$

where $P$ is the pressure, which is a function of the phase mix as well as $\rho$.

Previous studies suggest that at a given pressure there exists an equilibrium phase mix. Begelman & McKee (1990) derived an equilibrium equation of state as a function of pressure by considering the role of conduction for size scales larger than the Field length on the phase mix, and Elmegreen (1993) considered the equilibrium molecular to atomic gas fraction as a function of the pressure and radiation field. If we assume that the medium requires a relaxation timescale $\tau$ to achieve equilibrium, which we expect is greater than the cooling timescale in the individual phases, then we can describe the evolution towards equilibrium as a relaxation process. For $f$, the mass fraction of one phase of a two phase medium, we can describe the evolution towards equilibrium as

$$\frac{\partial(\rho f)}{\partial t} + \nabla \cdot (\rho f \mathbf{v}) = \frac{\rho(f_{eqm} - f)}{\tau} \tag{2.2.3}$$

where $\tau$ is the timescale required to obtain equilibrium and we have written the equilibrium phase fraction as $f_{eqm}$. We expect $f_{eqm}$ to be a function of local variables such as the pressure and the radiation field. We note that relaxation terms can strongly effect the dynamics of the Euler equation and can produce dissipation like terms in this equation (Whitham 1974 in traffic flow and molecular gases, Glimm 1986 in multiphase fluids, Liu 1991, Landau & Lifshitz 1959 in molecular gases, see bulk viscosity).

### 2.3 Sound Speeds and Critical Mach Number

We note that our set of equations is similar to the equations that describe relaxation effects in gases, where the internal energy may lag behind the equilibrium value. For example vibrational and rotational degrees of freedom may take longer to reach equilibrium than kinetic or translational ones in a molecular gas (see references cited above). In this section we follow the intuitive discussion by Whitham (1974) to derive a criterion for when shocks are continuous rather than sharp followed by a region of relaxation.

We can view the set of previous equations in terms of two systems. If we consider timescales of order $\tau$ then we require the full set of equations listed above; we call this system shock system I (Whitham's language). The characteristics for this system are $v, v \pm c$ for $v$ the velocity and $c$ the sound speed where

$$c^2 = \frac{\partial P(\rho, f)}{\partial \rho}. \qquad (2.3.1)$$

If we average the fluid flow over timescales larger than $\tau$ then we can consider the phase fraction to be at the equilibrium value. This is equivalent to the limit $\tau \to 0$ and is described by only two equations (neglecting the equation for the phase fraction), however the phase fraction is then equal at all times to the equilibrium value and the characteristics are $v \pm c'$, with sound speed $c'$ where

$$c'^2 = c^2 - d^2 \qquad (2.3.2)$$

and

$$d^2 = \frac{-\frac{\partial P}{\partial f} \frac{df_{eqm}}{dP} \frac{\partial P}{\partial \rho}}{1 - \frac{df_{eqm}}{dP} \frac{\partial P}{\partial f}}. \qquad (2.3.3)$$

In general we expect $\frac{\partial P}{\partial \rho} > 0$. For $f$ representing the denser phase fraction we expect that at higher pressure there should be a higher fraction of dense gas so that $\frac{df_{eqm}}{dP} > 0$. Similarly we expect $\frac{\partial P}{\partial f} < 0$ when the averaged density, $\rho$, is kept constant. Therefore for reasonable functions $d^2 > 0$ and

$$v - c < v - c' < v < v + c' < v + c \qquad (2.3.4)$$

which is a general condition for stability, in other words if the fluid begins with a smooth initial condition, discontinuities will not develop.

Viewed from the system II (averaging over timescales larger than $\tau$) which takes the flow between two uniform states in equilibrium, the shock will be continuous in system I if

$$v_2 - c_2 < U < u_1 + c_1 \qquad (2.3.5)$$

for $U$ the velocity of the shock. Subscripts here refer to quantities before and after the shock. A frozen shock followed by a region of relaxation will occur when

$$U > u_1 + c_1 \qquad (2.3.6)$$

or when the Mach number $M$

$$M \equiv \frac{U - u_1}{c_1'} > \frac{c}{c'}. \tag{2.3.7}$$

The critical Mach number $M_{crit} \equiv c/c'$.

## 2.4 Dispersion Relation, and Dissipative Behavior

Sometimes a dispersion relation can be used to gain some insight into the stability and dissipational behavior of a set of dynamical equations. Here we derive the dispersion relation for our set of equations and show for small $\tau$ that the flow is stable and dispersive. We derive the dispersion relation by considering the first order perturbations $\rho = \rho_0 + \epsilon \rho_a$, $v = v_0 + \epsilon v_a$, and $f = f_0 + \epsilon f_a$ where $\rho_a, v_a, f_a, \propto e^{ikx - i\omega t}$ and $\epsilon$ is small. We obtain the following dispersion relation

$$\left(-\omega^2 + 2kv_0\omega + k^2\left(c^2 + v_0^2\right)\right)\left(-i\omega + ikv_0 + \frac{1}{\tau} - \frac{1}{\tau}\frac{\partial f_{eqm}}{\partial f}\right) + \frac{k^2}{\tau}\frac{\partial P}{\partial f}\frac{\partial f_{eqm}}{\partial \rho} = 0 \tag{2.4.1}$$

for $c$ given above. In the limit $\tau \to 0$, $\omega = kv_0 \pm kc'$ and in the limit $\tau \to \infty$, $\omega = kv_0 \pm kc$. The solutions are stable and dispersive when all roots of the dispersion equation have $Im(\omega) < 0$. For small $\tau$ we can expand the dispersion relation in powers of $\tau$ and to first order in $\tau$ we find that

$$\omega = kv_0 \pm kc + \frac{ik^2\tau}{2}\frac{\partial P}{\partial f}\frac{df_{eqm}}{dP}\frac{\partial P}{\partial \rho}. \tag{2.4.2}$$

As described in the previous section, we we expect $\frac{\partial P}{\partial \rho} > 0$, $\frac{df_{eqm}}{dP} > 0$ and $\frac{\partial P}{\partial f} < 0$ so that the imaginary component of $\omega$ should be less than zero, and the solutions are damped. If we incorporate a kinematic viscosity $\nu$ into equation (2.2.2) and neglect phase changes, then we find to first order in $\nu$ that

$$\omega = kv_0 \pm kc - \frac{ik^2\nu}{2}. \tag{2.4.3}$$

It is clear from comparison of these two previous equations that the relaxation term in equation (2.2.3) behaves for small $\tau$ similar to a viscosity. If the ratio of the dissipation terms

$$\left(\frac{\tau}{\nu}\right)\frac{\partial P}{\partial f}\frac{df_{eqm}}{dP}\frac{\partial P}{\partial \rho} \gg 1 \tag{2.4.4}$$

then the dominant form of dissipation is due to phase changes for a system that has both phase changes and viscosity. This inequality should be roughly equivalent (in order of magnitude) to

$$\frac{\tau c^2}{\nu} \gg 1 \tag{2.4.5}$$

which we expect is satisfied when $\tau$ is greater than the mean collision time of clouds.

## 3. One-Dimensional Simulations

In this section we describe the results of a set of one-dimensional simulations of the equations given in §2.2. We run a one-dimensional computer code that is a 3rd order (in $L_1$ norm) finite difference scheme in conservative form which uses an ENO–LLF (Essential Non-Oscillatory, Local Lax-Friedrichs) shock capturing scheme (Shu & Osher 1988, 1989). The time stepping method is a 3rd order TVD (total variation bounded) Runge Kutta method. For a description of the implementation see Don & Quillen (1995) and Quillen (1995). Only the left hand side of equation (2.2.3) can be put in conservative form. We therefore treat the right hand side of this equation as a forcing term. Equations (2.2.2) and (2.2.1) are easily put in conservative form. Our simulations are run with 128 grid points with periodic boundary conditions in a two phase model. As described in the appendix we assume that the pressure is proportional to a power of the density in each phase (see equation A.5 for the form and notation). The denser phase is described with parameters $A_1$, $\mu_1$ and $\alpha_1$, whereas the less dense phase is described with parameters $A_2$, $\mu_2$ and $\alpha_2$. We use a model where $\alpha_1 = \alpha_2 = \alpha$. We assume an equilibrium phase fraction function (e. g. Elmegreen 1993)

$$f_{eqm}(P) = BP^\beta \qquad (3.1)$$

but don't allow $f_{eqm}$ to be greater than 1. Parameters for the simulations are listed in Table 1 and are unitless. The simulations run with grid spacing $dx = 1$ and are updated at a time interval of $dt = 0.2$.

We run a series of simulations with different values of $\tau$ and initial Mach number. Initial conditions are listed in Table 2 for the individual simulations which are denoted $S_a$, $S_b$, $T_a$ and $T_b$. Each simulation begins with a shock velocity that is zero in our inertial frame, and a phase fraction on either side of the shock that is at the equilibrium value. Initial conditions are a step function with initial values $\rho_l$, $f_l$ and $v_l$ one the left side of the interval and $\rho_r$, $f_r$ and $v_r$ one the right side of the interval. Because of the periodic boundary conditions the shock at the edge of the interval is a rarefraction wave.

In Figure 1 we present two simulations (denoted $S_a$ and $S_b$) with Mach number $M$ below the critical Mach number (see §2.3) that have different relaxation times $\tau$. $S_a$ has $\tau = 0.5$ and $S_b$ has $\tau = 5.5$. Note that the simulation with larger $\tau$ ($S_b$) has smoother shocks. This is because of the dissipative effect of the relaxation term in equation (2.2.3).

## 4. SUMMARY AND DISCUSSION

By incorporating phase changes in a simplistic way into a continuum fluid equation, we find that phase changes can significantly effect the gas dynamics of the ISM. The sound speed for the system with phase equilibrium should be lower than the sound speed of a system with a frozen phase mix. For small values of $\tau$ the timescale required to obtain an equilibrium phase mix, phase changes act similar to a viscosity. Simulated one-dimensional shocks with larger values of $\tau$ are smoother than those with smaller values of $\tau$, consistent with the dissipative nature of the system.

Many physical processes have not been considered in this paper. The role of radiation which can affect the phase mix and possibly cause star formation mediated radiative precursors may also strongly affect the continuum fluid dynamics. Turbulence and magnetic fields, also not considered here, could dominate the pressure (in the Euler's equation). In the centers of galaxies where rotation timescales are faster, (and in the outer parts of galaxies at small scales) departure from thermal equilibrium may occur. At small scales the effects of individual clouds cause the continuum fluid approximation to be invalid. It has been suggested (e. g. Quillen et al. 1995) that different phases in the ISM may not be moving together. In future more complicated sets of dynamical equations which can incorporate these processes should be studied.

It has recently become possible to observe the ISM in galaxies in a variety of wavelengths (and so phases); for example, it may become possible to constrain the radiation field and the cooling rate of the cooler components of the ISM as a function of position in a galaxy using far infrared observations. We hope that multi-wavelength observations of galaxies are coupled with 2-dimensional (and eventually 3-dimensional) fluid simulations of these galaxies, which can be run in a variety of fluid systems. This type of study will make it possible to constrain the fluid properties of the ISM.

We note that in barred and spiral galaxies the process of dissipation may determine the shape and location of the shocks as well as the gas inflow rate (Roberts, Shu & van Albada 1979, Barnes & Hernquist 1991). Whereas galactic disks are rather inefficient accretion disks (Steiman-Cameron & Durisen 1988) when shocks are formed in a non-axisymmetric gravitational potential mass transfer can occur rapidly (e. g. Mihos & Hernquist 1994a,b, Shlosman & Noguchi 1993). One possible explanation for this is that although the sheer viscosity may be small, the bulk viscosity (which becomes important in shocks and is present in numerical simulations) may be large. We note that when relaxation effects are important, it is primarily the bulk viscosity that is increased (e. g. Landau & Lifshitz 1959). In the barred galaxy NGC 7479, Quillen et al. (1995) have measured a gas inflow rate along the bar. It is likely that gas inflow in barred galaxies is strongly dependent upon the nature of the fluid (ISM). We hope in future to explore how gas inflow rates may depend upon the constituents of the ISM.

We acknowledge helpful discussions and correspondence with C. Thompson, B. Elmegreen, R. Olling, G. Shaviv, O. Regev, N. Shaviv, J. Glimm, A. Gould, K. Griest, P. Schecter, A. Goodman & D. DePoy. A.C.Q. acknowledges the support of a Columbus fellowship.

## Appendix. Equation of State, an Example

We can write the average density $\rho$ summed over a scale greater than the scale of individual clouds as

$$\rho = \sum_i f_{v_i} \rho_i \tag{A.1}$$

$$\rho^{-1} = \sum_i f_{m_i} \rho_i^{-1} \tag{A.2}$$

$$f_{v_i} = \frac{f_{m_i} \rho}{\rho_i} \tag{A.3}$$

where $f_{m_i}$, $f_{v_i}$ and $\rho_i$ are the mass fraction, volume filling fraction and density of phase $i$ respectively. We can also write

$$\rho = \sum_i \sigma_i \tag{A.4}$$

for $\sigma_i = f_{m_i}\rho$ the mass per unit volume of phase $i$ averaged over a volume which contains all phases. For an opticaly thin tracer of a particular gas phase the intensity observed would be proportional to $\sigma_i$ integrated along the line of sight.

The two phase model assumes that the pressure in all phases are the same. Studies of heating and cooling balance find that the pressure $P$ and $n_H$ the number density of $H$ atoms are related in each phase by scaling laws (Shull & Woods 1985) that look like

$$P_{gas} \approx A_i n_H^{\alpha_i} \tag{A.5}$$

for various assumptions about the UV, X-ray and cosmic ray fluxes. Using this we can write the density in each phase as

$$\rho_i = n_H \mu_i = (P/A_i)^{1/\alpha_i} \mu_i \tag{A.6}$$

for a mean molecular weight in phase $i$ of $\mu_i$. These equations provide enough information to solve for an "equation of state", $P(\rho, f_{m_i})$. This is is easiest to do if $\alpha$ is the same for all phases, in which case

$$P(\rho, f_{m_i}) = \rho^\alpha \left[ \sum_i f_{m_i} \mu_i^{-1} A_i^{1/\alpha} \right]^\alpha . \tag{A.7}$$

The derivatives $\frac{\partial P}{\partial \rho}$ and $\frac{\partial P}{\partial f_{m_i}}$ are needed to do simulations and estimate sound speeds, and these derivatives can be written

$$\frac{\partial P}{\partial \rho} = \left( \frac{P}{\rho} \right) \left[ \sum_i \frac{f_{v_i}}{\alpha_i} \right]^{-1} \tag{A.8}$$

$$\frac{\partial P}{\partial f_{m_i}} = P \left( \frac{\rho}{\rho_i} \right) \left[ \sum_j \frac{f_{v_j}}{\alpha_j} \right]^{-1} . \tag{A.9}$$

We note that the molecular gas is normally not included in the "three phase model of the ISM" (McKee 1990) because most molecular clouds are self-gravitating (Larson 1981, see McKee's review) and so are not in pressure equilibrium with the rest of the ISM. However, if there is a molecular to atomic phase transition, the pressure is indirectly dependent upon the mass fraction in the molecular phase. As Begelman & McKee (1990) have noted, physical processes causing phase changes such as conductivity and cloud evaporation can significantly effect the mean pressure and energy density of the medium. We can incorporate molecular clouds into the above model by considering their integrated density as a function of the pressure at the surface of the clouds. Elmegreen (1993) by considering statistics of self gravitating and diffuse molecular clouds, and self shielding of these clouds in a background UV radiation field, finds that the equilibrium molecular to atomic gas fraction is proportional to a power of the pressure.

TABLE 1

Parameters for One-D Simulations

| Parameter | |
|---|---|
| $A_1$ | 1.0 |
| $A_2$ | 2.0 |
| $\mu_1$ | 1.0 |
| $\mu_2$ | 1.0 |
| $\alpha$ | 1.3 |
| $B$ | 0.4 |
| $\beta$ | 1.0 |

TABLE 2

Initial Conditions
for One-D Simulations [1]

| Parameter | $S_a, S_b$ | $T_a, T_b$ |
|---|---|---|
| $M$ | 1.13 | |
| $M_{crit}$ | 1.31 | |
| $\rho_l$ | 1.32 | |
| $\rho_r$ | 0.86 | |
| $v_l$ | 0.85 | |
| $v_r$ | 1.29 | |
| $f_l$ | 0.80 | |
| $f_r$ | 0.60 | |

[1] $M$ is the initial Mach number and $M_{crit}$ is the critical Mach number (see equation 2.3.7). $\rho_l$, $v_l$ and $f_l$ are the initial density, velocity and mass fraction of the dense phase on the left hand side of the shock. $\rho_r$, $v_r$ and $f_r$ are the initial density, velocity and mass fraction of the dense phase on the right hand side of the shock.

# FIGURE CAPTIONS

**Figure 1.** Comparison of two $M < M_{crit}$ shocks with different values of $\tau$, the relaxation time. Parameters are listed in Tables 1 and 2. The solid line is simulation $S_a$ and the dotted line is simulation $S_b$. The two simulations have identical initial conditions except that $S_a$ has relaxation time $\tau = 0.5$ and $S_b$ has relaxation time $\tau = 5.5$. The simulations are shown at a time of $t = 11.8$. Note that the one with larger $\tau$ has a smoother profile, consistent with the dissipative behavior caused by the phase changes.

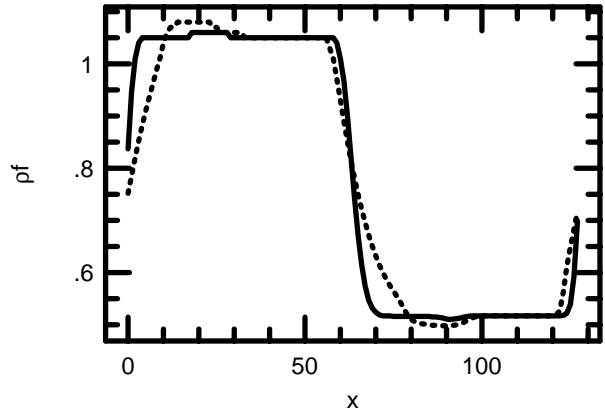
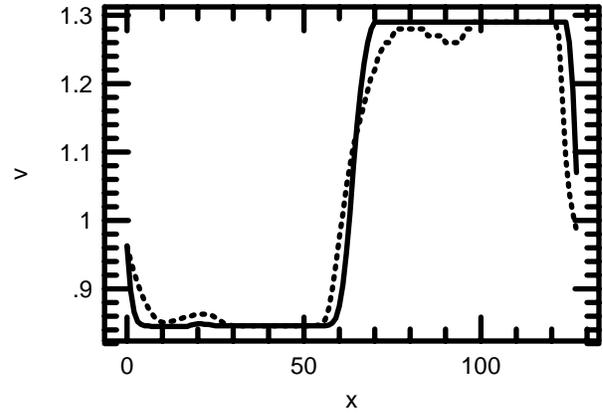
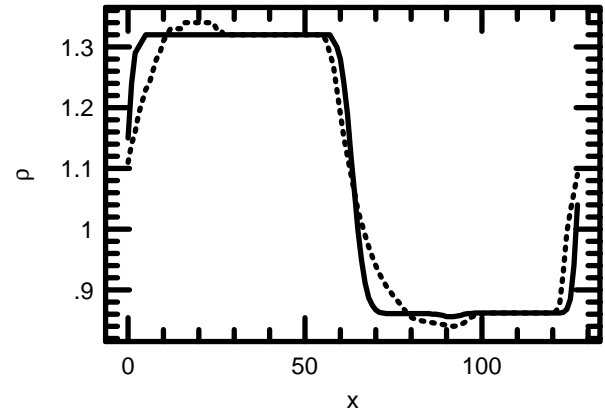